\RequirePackage{fixltx2e}

\documentclass[aps,prc,reprint,superscriptaddress]{revtex4-1}

\usepackage{graphicx}
\usepackage[caption=false]{subfig}
\graphicspath{{./graphics/}}  
\usepackage{amsmath}
\usepackage{array}
\newcolumntype{+}{>{\global\let\currentrowstyle\relax}}
\newcolumntype{^}{>{\currentrowstyle}}

\usepackage{scrextend}
\usepackage{hyperref}
\newcommand{\comment}[1]{}

\newcommand{\degree}{\ensuremath{^\circ}}
\usepackage{rotating}          

\newcommand{\gtaeq}{\lower 2pt \hbox{$\, \buildrel {\scriptstyle >}\over {\scriptstyle \sim}\,$}}
\newcommand{\ltaeq}{\lower 2pt \hbox{$\, \buildrel {\scriptstyle <}\over {\scriptstyle \sim}\,$}}

\usepackage[export]{adjustbox} 

\begin{document}




\title{Beta-delayed-neutron studies of \textsuperscript{135,136}Sb and \textsuperscript{140}I performed with trapped ions}


\author{B.~S.~Alan}
\email[]{alan2@llnl.gov}
\affiliation{Lawrence Livermore National Laboratory, Livermore, California 94550, USA}
\affiliation{Department of Nuclear Engineering, University of California, Berkeley, California 94720, USA}

\author{S.~A.~Caldwell}
\affiliation{Department of Physics, University of Chicago, Chicago, Illinois 60637, USA}
\affiliation{Physics Division, Argonne National Laboratory, Lemont, Illinois 60439, USA}

\author{N.~D.~Scielzo}
\affiliation{Lawrence Livermore National Laboratory, Livermore, California 94550, USA}

\author{A.~Czeszumska}
\affiliation{Department of Nuclear Engineering, University of California, Berkeley, California 94720, USA}

\author{J.~A.~Clark}
\affiliation{Physics Division, Argonne National Laboratory, Lemont, Illinois 60439, USA}
\affiliation{Department of Physics and Astronomy, University of Manitoba, Winnipeg, Manitoba R3T 2N2, Canada}

\author{G.~Savard}
\affiliation{Physics Division, Argonne National Laboratory, Lemont, Illinois 60439, USA}
\affiliation{Department of Physics, University of Chicago, Chicago, Illinois 60637, USA}

\author{A.~Aprahamian}
\affiliation{Department of Physics, University of Notre Dame, Notre Dame, Indiana 46556, USA}

\author{M.~T.~Burkey}
\affiliation{Department of Physics, University of Chicago, Chicago, Illinois 60637, USA}
\affiliation{Physics Division, Argonne National Laboratory, Lemont, Illinois 60439, USA}

\author{C.~J.~Chiara}
\affiliation{Physics Division, Argonne National Laboratory, Lemont, Illinois 60439, USA}
\affiliation{Department of Chemistry and Biochemistry, University of Maryland, College Park, Maryland 20742, USA}

\author{J.~Harker}
\affiliation{Physics Division, Argonne National Laboratory, Lemont, Illinois 60439, USA}
\affiliation{Department of Chemistry and Biochemistry, University of Maryland, College Park, Maryland 20742, USA}

\author{A.~F.~Levand}
\affiliation{Physics Division, Argonne National Laboratory, Lemont, Illinois 60439, USA}

\author{S.~T.~Marley}
\affiliation{Department of Physics, Louisiana State University, Baton Rouge, Louisiana 70803, USA}
\affiliation{Department of Physics, University of Notre Dame, Notre Dame, Indiana 46556, USA}

\author{G.~E.~Morgan}
\affiliation{Department of Physics and Astronomy, University of Manitoba, Winnipeg, Manitoba R3T 2N2, Canada}
\affiliation{Physics Division, Argonne National Laboratory, Lemont, Illinois 60439, USA}

\author{J.~M.~Munson}
\affiliation{Department of Nuclear Engineering, University of California, Berkeley, California 94720, USA}

\author{E.~B.~Norman}
\affiliation{Department of Nuclear Engineering, University of California, Berkeley, California 94720, USA}

\author{A.~Nystrom}
\affiliation{Department of Physics, University of Notre Dame, Notre Dame, Indiana 46556, USA}
\affiliation{Physics Division, Argonne National Laboratory, Lemont, Illinois 60439, USA}

\author{R.~Orford}
\affiliation{Department of Physics, McGill University, Montr\'eal, Qu\'ebec H3A 2T8, Canada}
\affiliation{Physics Division, Argonne National Laboratory, Lemont, Illinois 60439, USA}

\author{S.~W.~Padgett}
\affiliation{Lawrence Livermore National Laboratory, Livermore, California 94550, USA}

\author{A.~P\'erez Galv\'an}
\affiliation{Physics Division, Argonne National Laboratory, Lemont, Illinois 60439, USA}
\affiliation{Vertex Pharmaceuticals, San Diego, California 92121, USA}

\author{K.~S.~Sharma}
\affiliation{Department of Physics and Astronomy, University of Manitoba, Winnipeg, Manitoba R3T 2N2, Canada}

\author{K.~Siegl}
\affiliation{Department of Physics, University of Notre Dame, Notre Dame, Indiana 46556, USA}

\author{S.~Y.~Strauss}
\affiliation{Department of Physics, University of Notre Dame, Notre Dame, Indiana 46556, USA}


\date{\today}

\begin{abstract}
Beta-delayed-neutron ($\beta$n) spectroscopy was performed using the Beta-decay Paul
Trap and an array of radiation detectors.
The $\beta$n branching ratios and energy spectra for \textsuperscript{135,136}Sb and
\textsuperscript{140}I were obtained by measuring the time of flight of recoil ions
emerging from the trapped ion cloud.
These nuclei are located at the edge of an isotopic region identified as having 
$\beta$n branching ratios that impact the \textit{r}-process abundance pattern around
the $A$$\sim$$130$ peak. 
For \textsuperscript{135,136}Sb and \textsuperscript{140}I, $\beta$n branching ratios
of 14.6(11)$\%$, 17.6(28)$\%$, and 7.6(28)$\%$ were determined, respectively.
The $\beta$n energy spectra obtained for \textsuperscript{135}Sb and \textsuperscript{140}I
are compared with results from direct neutron measurements, and the $\beta$n energy
spectrum for \textsuperscript{136}Sb has been measured for the first time.

\end{abstract}


\maketitle{}


\section{Introduction}
\label{sec:Introduction}

Beta-delayed-neutron ($\beta$n) emission is a process that can occur
for neutron-rich nuclei sufficiently far from stability.  
In this process, a precursor nucleus undergoes $\beta^{-}$ decay to a highly excited
state in the daughter nucleus above the neutron-separation energy that emits a neutron.  
The properties of $\beta$n-emitting nuclei are important in various areas of basic and
applied sciences, including nuclear astrophysics, nuclear energy, and nuclear structure. 

The astrophysical rapid neutron-capture process (\textit{r}~process) is
believed to be responsible for the production of roughly half of the
elements heavier than iron~\cite{Burbidge1957,Cameron1957}.
In the \textit{r}~process, neutron-rich nuclei far from stability are produced through
repeated neutron-capture reactions, and $\beta$n emission during the eventual decay
back to stability impacts the final isotopic abundance pattern.
Different astrophysical environments, such as core-collapse supernovae~\cite{Meyer1992,Farouqi2010}
and neutron-star mergers~\cite{Goriely2011,Pian2017}, have been investigated as possible \textit{r}-process
sites by comparing theoretical models with observation.
These models require high-quality nuclear data, such as nuclear masses,
$\beta$-decay and neutron-capture rates, and $\beta$n-emission probabilities, for
the thousands of isotopes along the nucleosynthesis pathway and populated during the
decay back to stability.
Much of this information still remains unknown, given the
experimental challenges of accessing nuclei far from stability.

Beta-delayed-neutron emission also plays a key role in the control and safety of nuclear reactors.
Both the branching ratios and energy spectra are required for reactor kinetics
calculations and safety studies~\cite{Das1994,DAngelo2002}. 
Higher-quality nuclear data would allow for the $\beta$n yield and energy spectrum to be
calculated for individual contributing isotopes, making it possible to accurately
model any fuel-cycle concept, actinide mix, or irradiation history.

In addition, the information obtained in $\beta$n measurements helps to provide a better understanding of the nuclear structure
of neutron-rich nuclei~\cite{Kratz1984,Raman1983,Hamilton1985,Winger2009}. 
For example, measuring the $\beta$n-emission probability can be used to deduce the
$\beta$-strength function above the neutron-separation energy of the daughter
nucleus~\cite{Pappas1972,Kratz1979}.
Beta-delayed-neutron studies also help to constrain nuclear-structure calculations~\cite{Kawano2008}
and empirical models~\cite{McCutchan2012} that predict the decay properties of nuclei
for which no data exist.

In this work, the Beta-decay Paul Trap (BPT)~\cite{Yee2013,Scielzo2012,Scielzo2014},
a linear radiofrequency quadrupole ion trap with an open geometry,
was utilized to study the $\beta$n branching ratios and energy spectra of a number of
$\beta$n-emitting nuclei, which 
were produced with the Californium Rare Isotope Breeder Upgrade (CARIBU) facility~\cite{Savard2008}
at Argonne National Laboratory.
The results for $^{137,138}$I and $^{144,145}$Cs are discussed in
Ref.~\cite{Czeszumska_Paper}, and the results for the more neutron-rich isotopes,
\textsuperscript{135,136}Sb and \textsuperscript{140}I, are discussed here.
Recent sensitivity studies performed by Mumpower \textit{et al.}~\cite{Mumpower2016} indicate
that the latter three nuclei are situated at the edge of a region in the nuclear chart
where the $\beta$n branching ratios significantly impact the final
\textit{r}-process abundance pattern around the $A$$\sim$$130$ peak.

\section{Experimental Methods}
\label{sec:ExperimentalMethods}
\begin{figure}[!htbp]
   \includegraphics[width=0.5\textwidth]{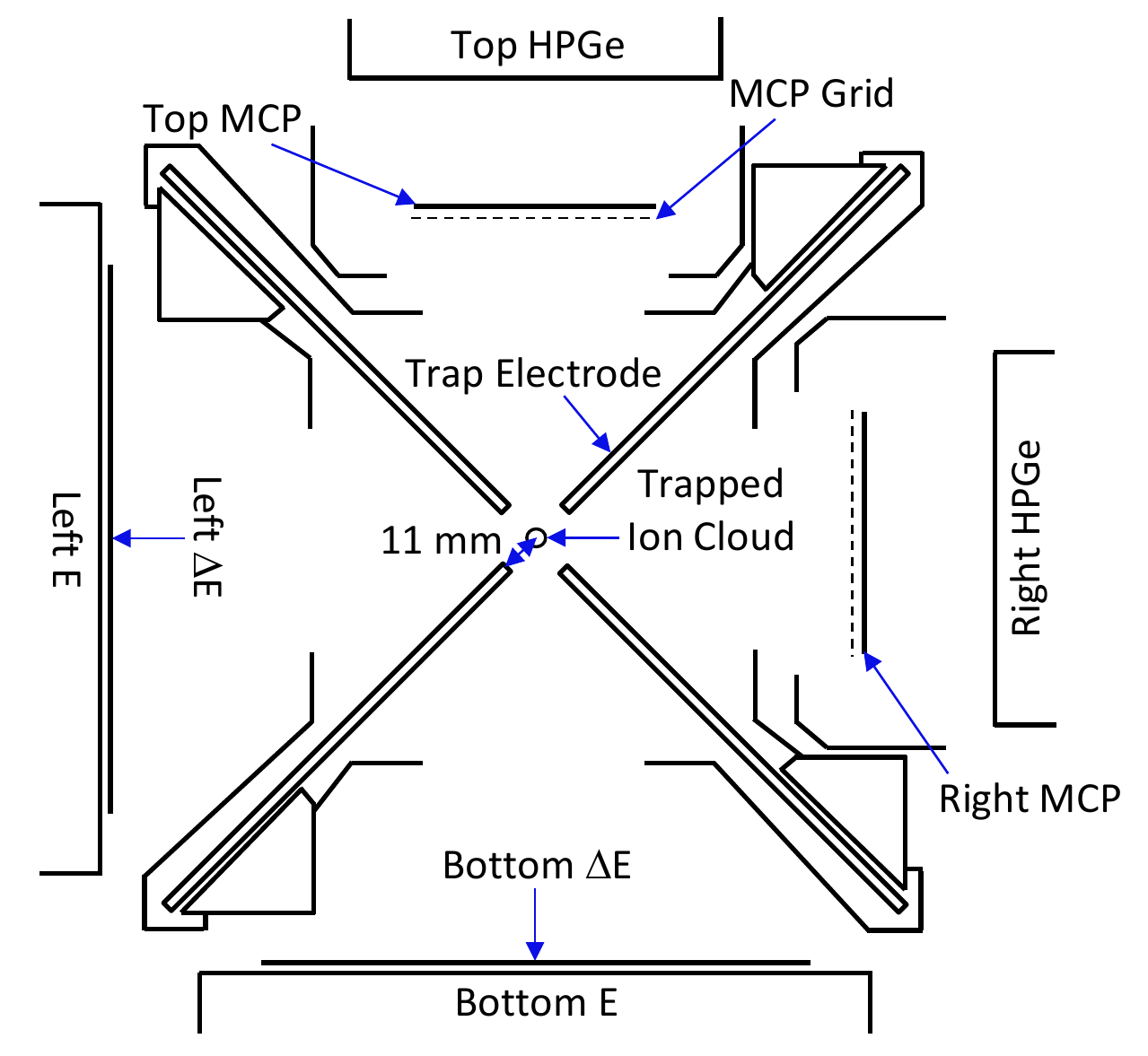}
   \caption
      {
      \label{fig:IonTrapSchematic}
      (Color online) Cross-sectional view of the BPT and detectors used in the 
      experiment (not to scale); the beam axis points perpendicularly into the plane.  
      The detectors are labeled by their orientation relative to the beam direction
      at the center of the trap.
      Two plastic $\Delta E$-$E$ telescopes, two MCP detectors,
      and two HPGe detectors were used to measure $\beta$ particles, recoil ions,
      and $\gamma$ rays, respectively. 
      Four sets of electrode plates were used to confine ions in the trap.  
      Each plate came within 11~mm of the center of the BPT. 
      }
\end{figure}

In the present work, the challenges associated with direct neutron detection
are circumvented by instead studying the nuclear recoil from $\beta$ decay. 
Radioactive ions are suspended in vacuum as a 
$\sim$1-mm\textsuperscript{3} cloud at the center of the BPT.  
When a trapped ion undergoes $\beta$ decay, the recoil ion and emitted radiation emerge from the 
cloud with negligible scattering, allowing for their properties to be measured with 
radiation detectors arranged around the BPT as shown in Fig.~\ref{fig:IonTrapSchematic}.
Two plastic-scintillator $\Delta E$-$E$ telescopes, two microchannel-plate (MCP) detectors,
and two high-purity germanium (HPGe) detectors are used to measure $\beta$ particles, recoil ions, and 
$\gamma$ rays, respectively. 

Beta-delayed-neutron spectroscopy is performed by recording the time of flight (TOF) of the
recoil ions, which is determined from the time difference between the $\beta$ particle
hitting a $\Delta E$ detector and the recoil ion hitting an MCP detector.  
Due to the additional momentum imparted by the neutron, ions from $\beta$n emission
have shorter TOFs than those from $\beta$ decay without neutron emission.
The recoil-ion momentum can be reconstructed from the TOF and the distance the ion travels
to the MCP surface.  The neutron energy may then be obtained through conservation of
energy and momentum. 
The resulting neutron-energy spectrum can be determined down to 100~keV;
at lower energies, TOF cannot be used to identify $\beta$n events because the corresponding
recoil ions have energies comparable to those from 
$\beta$ decays without neutron emission.
In this section, the ion production, transport, and confinement, as well as
the detection of the decay particles are discussed.

\subsection{Beam delivery at CARIBU}
\label{sec:BeamDeliveryAtCARIBU}

At CARIBU, fission fragments from a $\sim$100-mCi \textsuperscript{252}Cf source
were thermalized in a large helium-filled gas catcher~\cite{Savard2008}, extracted
primarily as $1^+$ ions, transported through an isobar separator~\cite{Davids2008}, and
delivered to a radiofrequency-quadrupole buncher containing a small amount of helium
gas to accumulate, cool, and bunch the beam. The isobar separator had a mass resolution
of $M/\Delta M \approx$ 14000, which allowed for
some suppression of isobars one neutron away from the desired species and
essentially complete removal of isobars more than one neutron away.

The optimal isobar-separator settings were selected by monitoring the distribution of
isotopes present in the beam during tuning.
The beam composition was characterized by using the two HPGe detectors surrounding the BPT
and by performing mass scans with the Canadian Penning Trap (CPT) mass
spectrometer~\cite{Savard1997, Wang2004}.
The ion bunches were injected into the BPT at time intervals of $t$\textsubscript{int} and
accumulated over a length of time $t$\textsubscript{meas},
after which the ions were ejected from the trap to measure backgrounds over a time period
$t$\textsubscript{bkgd}; this cycle was repeated throughout the entire run.
The values of $t$\textsubscript{int}, $t$\textsubscript{meas}, and $t$\textsubscript{bkgd}
used for each isotope are given in Table~\ref{tab:RunInformation}
and were chosen based on the radioactive half-life of the isotope being studied and the
distribution of isobaric contaminants present during the measurement.
The total measurement times and average beam rates are also shown in Table~\ref{tab:RunInformation}.

\begin{table*}
\caption{\label{tab:RunInformation}
	The measurement time, average beam rate, and trapping-cycle information
	($t$\textsubscript{int}, $t$\textsubscript{meas}, $t$\textsubscript{bkgd})
	for the measurements.
	During each measurement cycle, ion bunches were injected into the BPT at
	time intervals of $t$\textsubscript{int}, accumulated over a length of time
	$t$\textsubscript{meas}, then ejected from the BPT for a background measurement
	lasting $t$\textsubscript{bkgd}.
	} 
\begin{ruledtabular}
\begin{tabular}{ l c c c c c c }
  Isotope & Half-life & Measurement time & Average beam rate & $t$\textsubscript{int} & $t$\textsubscript{meas} & $t$\textsubscript{bkgd}\\ 
  & (s) & (h) & (ions/s) & (s) & (s) & (s)\\ 
  \hline\noalign{\smallskip}
  \textsuperscript{135}Sb & 1.679(15)~\cite{NuclearDataSheets_135} & 45.7 & 50 & 1.0 & 19.9 & 10.1\\ 
  \textsuperscript{136}Sb & 0.923(14)~\cite{NuclearDataSheets_136} & 60.7 & 5 & 0.6 & 8.9 & 4.9\\ 
  \textsuperscript{140}I & 0.86(4)~\cite{NuclearDataSheets_140} & 35.3 & 5 & 0.6 & 8.3 & 4.3\\ 
\end{tabular} 
\end{ruledtabular}
\end{table*}

\subsection{Trapping with the BPT}
\label{sec:TrappingWithTheBPT}

Ion confinement was achieved by applying direct-current (DC) and time-varying, 
sinusoidal radiofrequency (RF) voltages to four sets of electrode plates 
extending to within 11~mm from the center of the trap as shown in 
Fig.~\ref{fig:IonTrapSchematic}. 
The DC voltages were used to produce a harmonic confining potential 
with a $\sim$5-V electrostatic valley in the axial direction,  
and the RF voltages, with a peak-to-peak amplitude of
about 200~V and a frequency of 310~kHz, were used to confine 
ions in the radial direction.   
Higher harmonics at 620 and 930~kHz were observed with amplitudes
less than 10\% of the amplitude of the primary frequency.
The trapped ions were thermalized in $\sim$5$\times$$10^{-5}$~Torr
of helium gas.

Following $\beta$ decay, the charge state of the recoil ion is typically $2^+$;
however, higher charge states can arise due to processes such as
electron shakeoff, Auger-electron emission, and
internal conversion.
The stability conditions for the BPT, determined from the Mathieu equations~\cite{Paul1990},
were chosen so that the decay daughters, which all have charge states higher than $1^+$,
were not confined in the trap.

\subsection{Particle detection}
\label{sec:ParticleDetection}

Two plastic-scintillator $\Delta E$-$E$ telescopes were used for 
$\beta$ spectroscopy.  
The $\Delta E$ detector was a 1-mm-thick, 
10.6-cm-diameter disk that had a nearly 100\% intrinsic detection efficiency
for $\beta$ particles and only a $\sim$1\% intrinsic detection efficiency 
for $\gamma$ rays and neutrons. 
The $\Delta E$ detectors were placed $\sim$105~mm from the center of the BPT
and each covered a solid angle of 5$\%$ of $4\pi$.
The $E$ detectors were 10.2-cm-thick, 13.3-cm-diameter disks located immediately
behind the $\Delta E$ detectors that were capable of stopping the $\beta$ particles.
Each $\Delta E$-$E$ telescope was contained in its own vacuum chamber
(held below $10^{-3}$ Torr) and separated from the BPT vacuum by a
10-$\mu$m-thick aluminized-Kapton window.
The Left and Bottom $\Delta E$ detectors had $\beta$-energy thresholds
of 76(24)~keV and 62(30)~keV, respectively,
and a neutron detection threshold of 370(70)~keV~\cite{Czeszumska_Paper}.

Two 50.3~$\times$~50.3~mm$^2$ resistive-anode Chevron MCP detectors~\cite{Wiza1979} 
with 1-ns timing resolution and sub-mm position sensitivity 
were used for recoil-ion detection.
The front face of each detector was biased to approximately $-2.5$~kV to 
accelerate incoming ions and thereby provide a more uniform detection efficiency.
Each detector was placed 4.5~mm behind a grounded 89\%-transmission grid   
to help shield the detector from the RF fields of the BPT and to prevent
the recoil-ion trajectories from being affected by the MCP bias voltage
until they passed through the grid.  
The hit locations of the ions were reconstructed from the relative amounts of charge 
collected at the four corners of the anode~\cite{Lampton1979}.
The central 46~$\times$~46~mm$^2$ region of each MCP detector had the best position
resolution and was taken to be the fiducial area in the data analysis.
Each detector was located 53.0(5)~mm away from the trap center and subtended a solid-angle
of 5$\%$ of $4\pi$.

The intrinsic efficiencies of the MCP detectors were determined to be 33.3(15)\% and 29.3(14)\%
for the Right and Top detectors, respectively, from a detailed study
of the decays of trapped $^{134}$Sb ions held in the BPT~\cite{Siegl2018}.
The ion detection efficiencies also had to be corrected for additional loss of MCP pulses
to electronic thresholds~\cite{Czeszumska_Paper, Czeszumska_Thesis}. For the Right MCP detector,
this was a $<$3\% correction. However, the Top MCP detector had a lower gain,
resulting in a correction that ranged between $\sim$5--30\% (depending on the
impact energy of the ions) and showed some spatial dependence.

Two coaxial single-crystal p-type HPGe detectors were used to detect $\gamma$ rays.
The detectors, which had relative efficiencies of 80\% and 140\%, were located within
10~cm of the trapped-ion cloud behind the Right and Top MCP detectors, respectively.
Standard $\gamma$-ray point sources ($^{60}$Co, $^{133}$Ba, $^{137}$Cs, $^{152}$Eu)
with activities determined to within 1.5-2.5\% (at 1$\sigma$) were used to calibrate the
photopeak detection efficiencies.

The data-acquisition system was triggered on a signal from any
detector.  A 22-$\mu$s coincidence window was then opened, during which
the amplitude and timing of each detected event was recorded along with the
phase of the BPT RF voltage.  
The TOF for recoil ions was determined with a timing resolution of 3~ns FWHM. 
The nonparalyzable deadtime per event was 142~$\mu$s.

\section{Analysis and Results}
\label{sec:AnalysisAndResults}

\begin{figure}[!htbp]
   \includegraphics[width=0.5\textwidth]{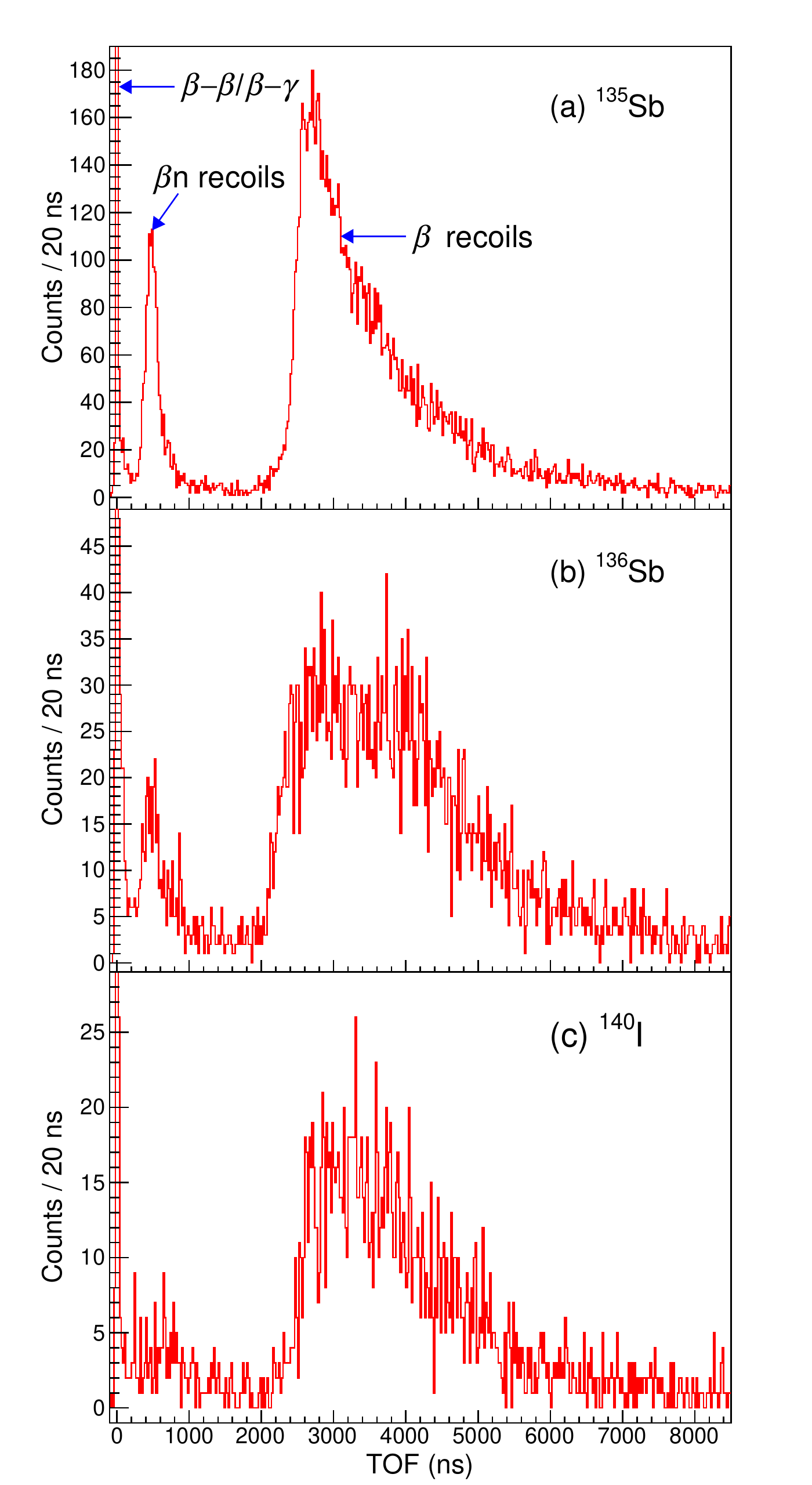}
   \caption
      {
      \label{fig:TOFSpectra}
      (Color online) TOF spectra for (a) \textsuperscript{135}Sb,
      (b) \textsuperscript{136}Sb, and (c) \textsuperscript{140}I.
      Events between 200 and 2000~ns are primarily due to recoil ions
      from $\beta$n decay, and events above 2000~ns are primarily
      due to recoil ions from $\beta$ decay without neutron emission.
      The peak at 0~ns is due to coincidences where a $\beta$ particle
      hit a $\Delta E$ detector and a $\gamma$ ray or scattered $\beta$
      particle triggered an MCP detector.  
      }
\end{figure}

The TOF of the recoil ions was determined from $\Delta E$-MCP detector coincidences and
used to distinguish $\beta$n decays from $\beta$ decays without neutron emission.
The TOF spectra measured for \textsuperscript{135,136}Sb and \textsuperscript{140}I
are shown in Fig.~\ref{fig:TOFSpectra}.
The $\beta$n events have TOFs primarily between 200 and 2000~ns, and 
$\beta$-decay events without neutron emission have longer TOFs.
A peak at 0~ns arose from $\beta$-particle events in the $\Delta E$ detector that were in
coincidence with a $\gamma$ ray or scattered $\beta$ particle triggering the
MCP detector. 

The $\beta$n energy spectra and branching ratios determined from these TOF spectra
are discussed in this section.
The Monte Carlo simulations of the decays and experimental setup
needed to analyze the data are introduced first.

\subsection{Monte Carlo simulations}
\label{sec:MonteCarloSimulations}

The $\beta$-decay kinematics were generated using simulation code originally
developed in Ref.~\cite{Scielzo2003} and later adapted for
$\beta$n decay~\cite{Yee2013,Siegl2018,Munson2018}. For each $\beta$-decay transition,
a distribution of $\beta$ and $\nu$ momenta was generated, assuming an allowed
$\beta$-spectrum shape. For excited states, the subsequent deexcitation to the
ground state by the emission of $\gamma$ rays, conversion electrons (CEs), and
neutrons was also included.  The resulting nuclear recoil was determined from
the momentum imparted from each of these decay particles.
For $\beta$n emission, the transitions were assumed to be allowed Gamow-Teller
and assigned a $\beta$-$\nu$ angular correlation, $a_{\beta\nu}$, of $-1/3$.
For transitions to states below the neutron-separation energy, an approximation
was made that for a given isotope, all the $a_{\beta\nu}$ were fixed to a single value, which
was determined from the measured $\beta$-ion coincidences using an approach described
in detail in Ref.~\cite{Munson2018}.  
For \textsuperscript{135}Sb and \textsuperscript{140}I, this value of $a_{\beta\nu}$
was $+0.23$ and $-0.42$, respectively.  For \textsuperscript{136}Sb, the presence of trapped
\textsuperscript{136}Te ions complicated the analysis of the recoil ions and
a value for $a_{\beta\nu}$ could not be obtained.

The $\beta$ decays were spatially distributed with a 1-mm-FWHM
Gaussian distribution in three dimensions, corresponding to the measured
ion-cloud extent~\cite{Siegl2018}.
The emitted $\beta$ particles, $\gamma$ rays, CEs, and neutrons were propagated using
GEometry ANd Tracking 4 (GEANT4)~\cite{Agostinelli2003, Allison2006}
version 4.10.0.p01 to model the scattering and energy loss of the particles within
the apparatus. The energies deposited in the $\Delta E$, $E$, and HPGe detectors were recorded,
and the electronic thresholds of the $\Delta E$ detectors were taken into account.             
Recoil ions of various charge states were propagated through 
the time-varying electric fields of the BPT using the SIMION 8.1~\cite{SIMION} ion-optics code.
The average charge states following the decay of \textsuperscript{135,136}Sb and
\textsuperscript{140}I were determined to be 2.20, 2.51, and 2.16, respectively~\cite{Munson2018},
from the RF-phase dependence of the measured $\beta$-ion coincidence rate using the
approaches described in Ref.~\cite{Siegl2018}.
For ions that struck an MCP detector, a threshold cut was applied~\cite{Czeszumska_Paper}
and the TOF, energy, velocity, and position at impact were recorded.

The efficiencies for detecting $\beta$ particles and $\beta$-ion coincidences were
determined using these simulations.  Fig.~\ref{fig:FastEfficiencyCorrection} shows the
$\beta$-ion-coincidence detection efficiency as a function of neutron energy for
\textsuperscript{135}Sb, with the product of the corresponding detector solid angles and
MCP-detector intrinsic efficiency divided out.
At the highest neutron energies, the coincidence-detection efficiency drops
rapidly because of the limited energy available for the leptons, which results
in fewer $\beta$ particles having energies above the $\Delta E$ detector thresholds.
However, $\beta$ decays that populate highly-excited states are largely suppressed
because of phase-space considerations.
The two $180\degree$ combinations (Left-Right and Bottom-Top) have higher
efficiencies than the two $90\degree$ combinations (Left-Top and Bottom-Right)
primarily because of neutron-ion coincidences, which are present because the neutron and
recoil ion are emitted with momenta nearly $180\degree$ apart and therefore strike back-to-back detectors. 
The $\beta$-ion-coincidence detection-efficiency curves for \textsuperscript{136}Sb
and \textsuperscript{140}I have similar features.

\begin{figure}[!htbp]
   \includegraphics[width=0.5\textwidth]{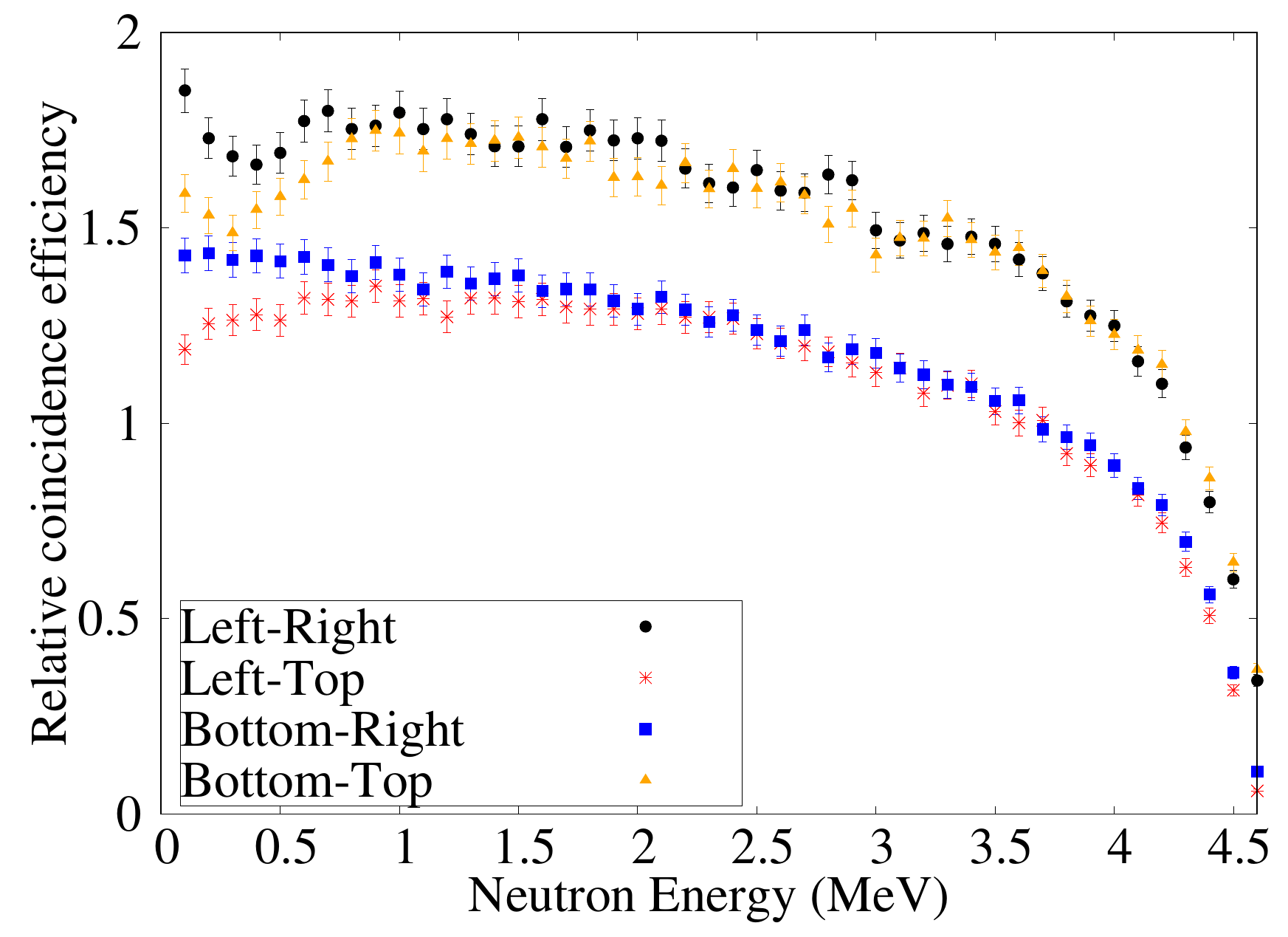}
   \caption
      {
      \label{fig:FastEfficiencyCorrection}
      (Color online) The $\beta$-ion-coincidence detection efficiency
      for each $\Delta E$-MCP detector pair as a function of neutron energy for
      \textsuperscript{135}Sb; the product of the corresponding detector solid angles
      and MCP-detector intrinsic efficiency has been divided out.
      The two $180\degree$ combinations (Left-Right and Bottom-Top) have higher
      efficiencies than the two $90\degree$ combinations (Left-Top and Bottom-Right)
      primarily because of additional events from neutron-ion coincidences. 
      At the highest neutron energies, the coincidence detection efficiency drops
      rapidly because of the limited energy available for the leptons;
      however, few $\beta$ decays are expected to yield neutrons at these
      energies because of phase-space considerations.
      The $\beta$-ion-coincidence detection-efficiency curves for \textsuperscript{136}Sb
      and \textsuperscript{140}I have similar features.
      }
\label{fig:FastEfficiencyCorrection}
\end{figure}

\subsection{Neutron energy spectra}
\label{sec:NeutronEnergySpectra}

The neutron energy was obtained by assuming the recoil ion and
neutron had equal and opposite momenta.  The momentum was approximated
from the recoil-ion TOF and hit position on the MCP surface; the distance
traveled by the recoil ion was approximated as a straight path from the trap center
to the MCP grid, and effects due to the acceleration of the ion between the
grid and the MCP surface were corrected for analytically.  Background events
were subtracted from the resulting spectrum, after which the neutron energy was adjusted
to account for the contribution to the recoil-ion momentum from lepton emission.
The spectrum was then scaled by the $\beta$-ion coincidence efficiency.
Each of these data analysis steps is explained below. 
 
The background from accidental coincidences was determined from the TOF region
between 15--20 $\mu$s, which both data and simulation indicated had no
true $\beta$-ion coincidences.  
This subtraction resulted in a 3--9$\%$ correction, depending on the isotope.

After accounting for accidental coincidences, counts remained in the
50--200-ns time window where no $\beta$-ion coincidences from trapped
ions were expected.
These counts were present both while the BPT was trapping ions and while the BPT was
held empty following ejection of the trapped ions and were likely due to radioactivity
that accumulated on the BPT and detector surfaces during data collection.
The TOF distribution of these events was most pronounced between 50--200 ns and
decreased with increasing TOF, extending into the $\beta$n TOF region.
The shape of this background, when converted into a neutron-energy distribution,
closely resembled an exponential function. 
The subtraction of this background was performed by normalizing this exponential
function to match the number of counts between 50-—200 ns collected when the BPT
was trapping ions. This resulted in a 15--30$\%$ correction to the total number of
observed $\beta$n decays, depending on the isotope being analyzed. 

During data collection, isobaric contaminants one neutron away from the isotope of interest
were suppressed but not completely removed.
For \textsuperscript{135,136}Sb and \textsuperscript{140}I, 
the more neutron-rich isobar (\textsuperscript{135,136}Sn and \textsuperscript{140}Te, respectively) 
is a $\beta$n emitter, but has a $^{252}$Cf-fission yield a couple orders of magnitude lower
than the isotope of interest, making its contribution to the total number of $\beta$n decays
in the BPT negligible. 
For \textsuperscript{135}Sb and \textsuperscript{140}I, the more proton-rich isobar
(\textsuperscript{135}Te and \textsuperscript{140}Xe, respectively) 
does not decay by $\beta$n emission and therefore cannot contribute $\beta$n events.
For \textsuperscript{136}Sb, the more proton-rich isobar, \textsuperscript{136}Te,
has a $\beta$n branching ratio roughly ten times smaller than that of \textsuperscript{136}Sb,
but a fission yield 30 times larger.
The suppression of \textsuperscript{136}Te by the isobar separator, together with
the measurement cycle favoring the shorter-lived species,
resulted in an average trapped-ion activity with about 15\% more \textsuperscript{136}Sb
than \textsuperscript{136}Te.
The \textsuperscript{136}Te contribution to the total number of $\beta$n coincidences was determined
to be 5\% based on the ratio of the \textsuperscript{136}Sb and \textsuperscript{136}Te activities,
after accounting for the $\beta$n branching ratios and
the fraction of neutrons with energies above the 100-keV neutron threshold
(estimated to be 0.6(2) for $^{136}$Te from the neutron-energy spectrum in Ref.~\cite{Shalev1974}
and determined in Sec.~\ref{sec:BDNBranchingRatios} to be 0.89(6) for $^{136}$Sb). 

The neutron energy determined solely from the recoil-ion momentum was adjusted to account for
the momentum imparted to the recoil ion from lepton emission.
For the $\beta$-ion coincidences measured by detectors $180\degree$ apart,
the neutron energy tended to be overestimated because the $\beta$ particle was
emitted in approximately the same direction as the neutron and therefore contributed
to the momentum of the nuclear recoil.
Simulations showed that neglecting the leptons resulted in an
overestimation of the inferred neutron energy of 25--30$\%$ at 100~keV, which steadily
decreased to 10$\%$, 7$\%$, and $<$4\% at neutron energies of 500~keV, 1000~keV,
and above 2000~keV, respectively.
For the $\beta$-ion coincidences measured by detectors $90\degree$ apart, 
the overestimation was only 1--2$\%$ for all neutron energies.
For \textsuperscript{136}Sb and \textsuperscript{140}I, the neutron emission was assumed
to directly populate the ground state, as there is currently no data indicating
that excited states are populated.
For \textsuperscript{135}Sb, however, decays to the ground state as well as the first, second,
and third excited states (populated with probabilities of 62$\%$, 21$\%$, 11$\%$, and 6$\%$,
respectively~\cite{Hoff1989}) were taken into account,
and the inclusion of the excited states led to a 1$\%$ shift in the inferred neutron energy.

\begin{figure}[!htbp]
   \subfloat
   {
     \includegraphics[width=0.5\textwidth]{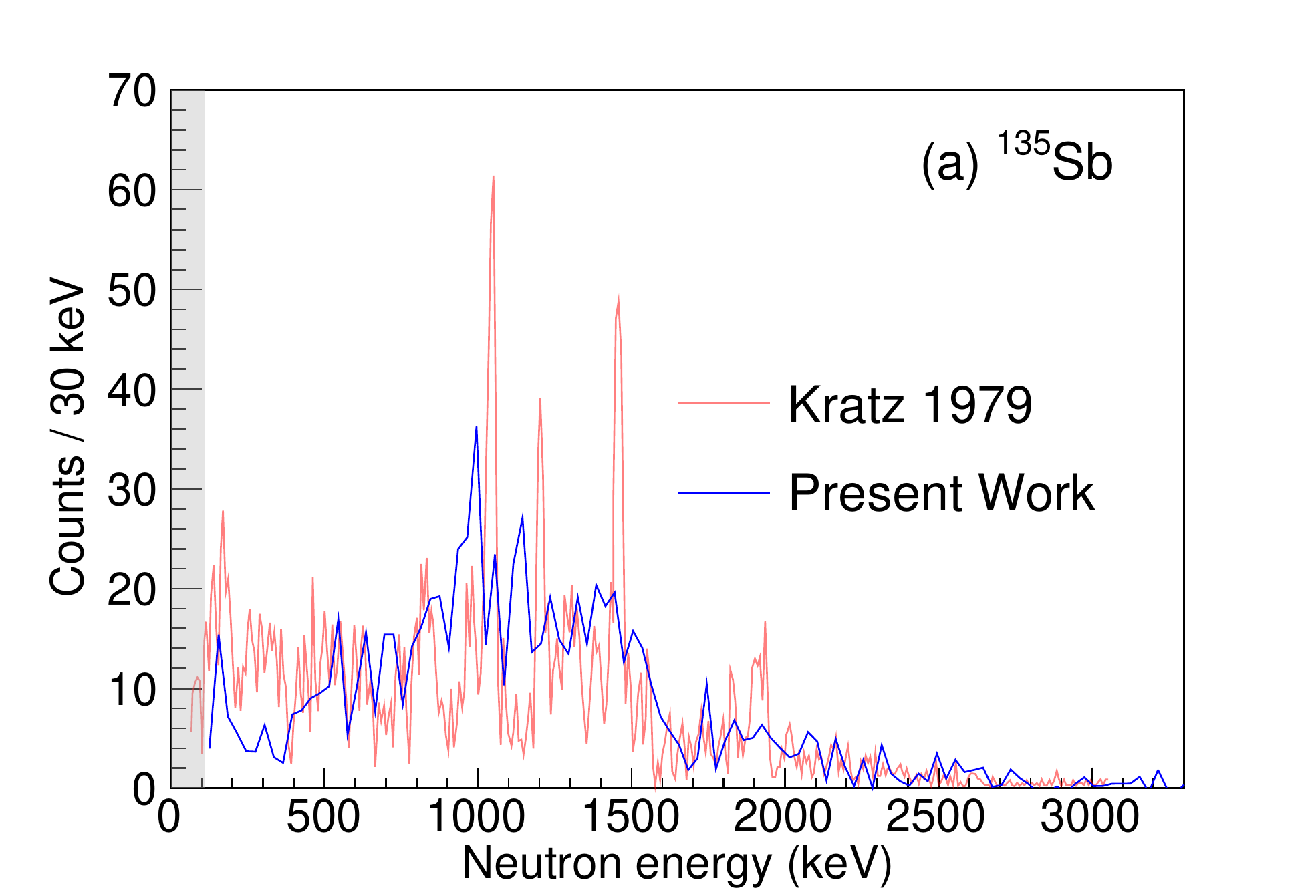}
   }
   \par
   \subfloat
   {
     \includegraphics[width=0.5\textwidth]{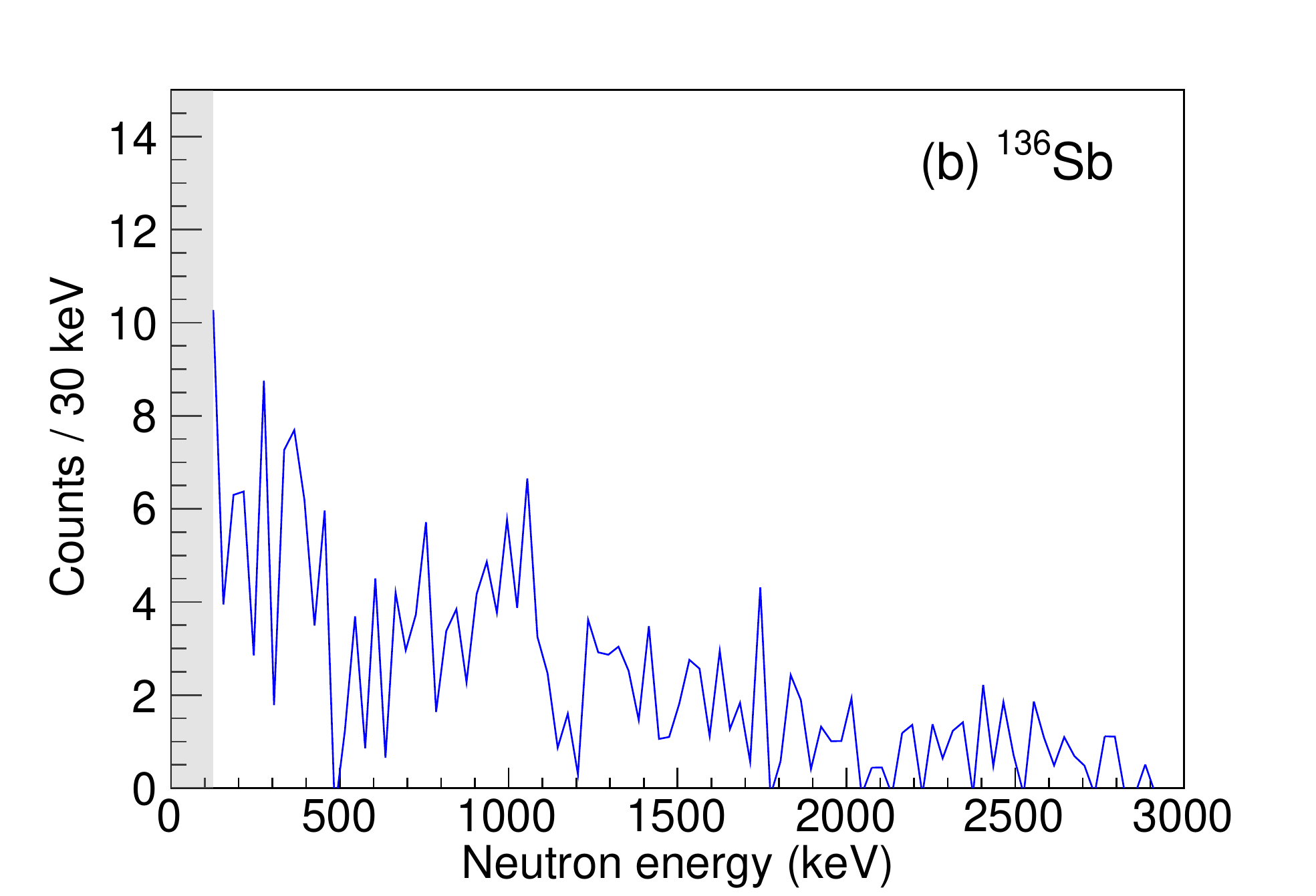}
   }
   \par
   \subfloat
   {
     \includegraphics[width=0.5\textwidth]{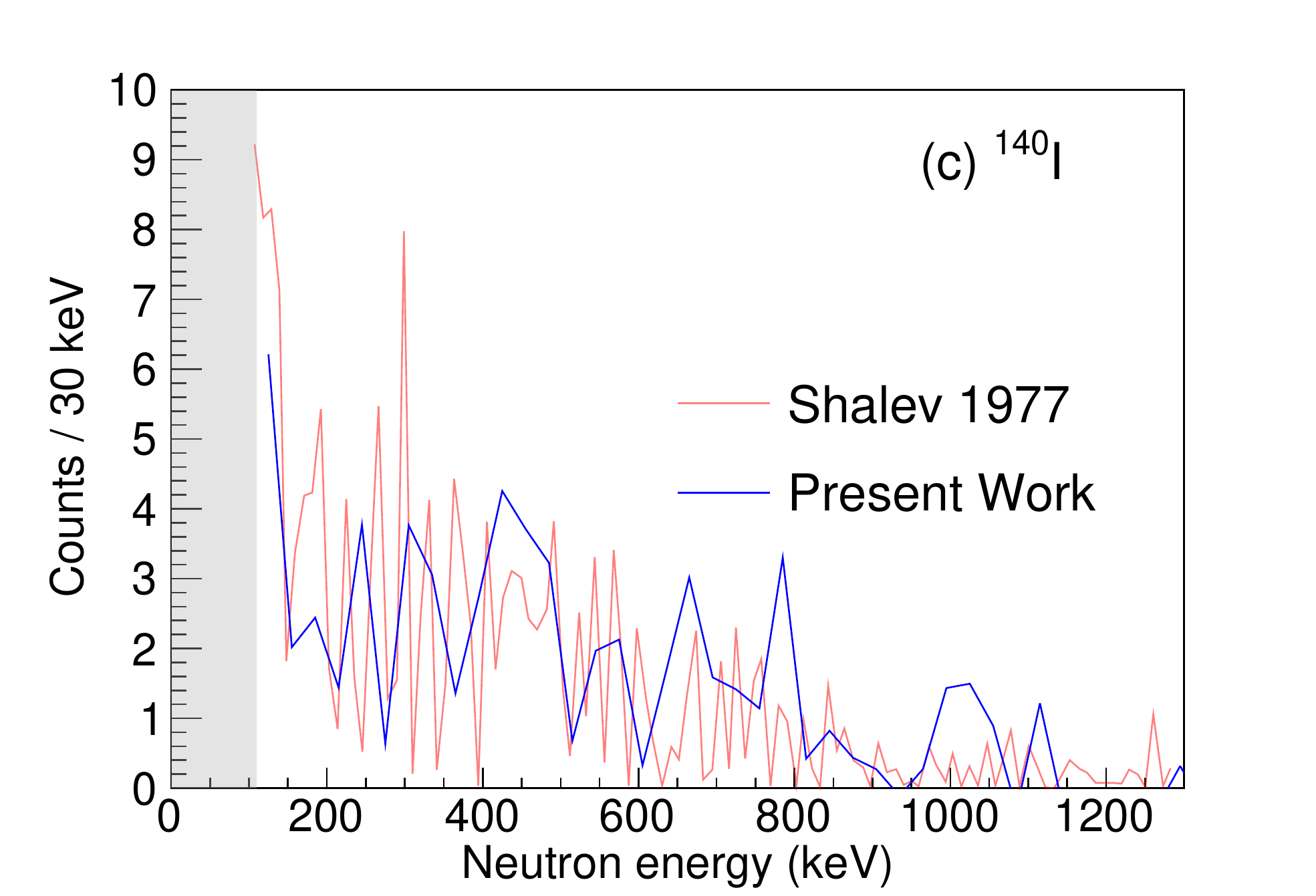}
   }
   \caption
      {
      \label{fig:NeutronEnergySpectra}
      (Color online) Neutron energy spectra for (a)~\textsuperscript{135}Sb,
      (b)~\textsuperscript{136}Sb, and (c)~\textsuperscript{140}I
      compared with results from Kratz~\textit{et al.}~\cite{Kratz1979}
      and Shalev and Rudstam~\cite{Shalev1977}.    
      The y-axis label refers to the present work, where each data point 
      corresponds to a 30-keV-wide bin.  For the $^{135}$Sb spectrum measured
      by Kratz~\textit{et al.} and the $^{140}$I spectrum measured by Shalev and Rudstam,
      the data points correspond to $\sim$8-keV-wide and $\sim$10.75-keV-wide bins,
      respectively.
      In the gray region below $\sim$100~keV, no neutron-energy information
      was obtained in the present work because the TOF of recoils from $\beta$n emission
      could not be distinguished from those from $\beta$ decay without neutron emission.
      }
\end{figure}

For each isotope, the neutron-energy spectrum obtained for each
$\Delta E$-MCP detector pair was corrected by the corresponding 
neutron-energy-dependent $\beta$-ion coincidence efficiency.
The results summed together for the four $\Delta E$-MCP detector pairs are shown in
Fig.~\ref{fig:NeutronEnergySpectra} for \textsuperscript{135,136}Sb
and \textsuperscript{140}I.
For $^{136}$Sb, the contribution from $^{136}$Te isobaric contamination was removed by
subtracting the $^{136}$Te neutron-energy spectrum measured in Ref.~\cite{Shalev1974},
which was scaled by the activity and $\beta$n branching ratio and broadened to account for
the experimental energy resolution.  The neutron-energy resolution in the present work
was primarily determined by the spatial distribution of the ion cloud and the spread in
recoil momentum resulting from the lepton emission.  Simulations indicated that the
FWHM energy resolution was 60$\%$ at a neutron energy of 100~keV and steadily decreased
to 25$\%$, 15$\%$, and 9$\%$ at 500~keV, 1000~keV, and above 2000~keV, respectively.
The neutron energy spectrum was determined down to $\sim$100~keV; below this energy,
the recoil momentum imparted from the emission of the leptons and any accompanying
$\gamma$ rays was comparable to the momentum imparted from neutron emission.

For \textsuperscript{135}Sb and \textsuperscript{140}I, the neutron-energy spectra are
compared with direct neutron measurements by Kratz~\textit{et al.}~\cite{Kratz1979}
and Shalev and Rudstam~\cite{Shalev1977}, respectively.
For \textsuperscript{136}Sb, no previous measurement of the energy spectrum has been made.
In the experiment by Kratz~\textit{et al.}, $\beta$n precursors were produced through
neutron-induced fission of \textsuperscript{235}U at the Mainz TRIGA reactor,
and two $^3$He ionization chambers, with energy resolutions of $\sim$12 keV for thermal
neutrons and $\sim$20 keV for 1-MeV neutrons, were used to measure neutron energies.
In the experiment by Shalev and Rudstam, $\beta$n precursors were produced at the
OSIRIS isotope-separator on-line facility.  Neutron energies were measured with a   
neutron spectrometer that consisted of a cylindrical gridded ionization chamber filled
with a $^3$He-argon gas mixture.
The results obtained with the BPT for \textsuperscript{135}Sb and \textsuperscript{140}I
have neutron-energy spectra and energy thresholds that are similar to the direct measurements.
For \textsuperscript{135}Sb, the peaks in the spectrum obtained here are not as sharp because of the
wider energy resolution; it also appears that there is a 4$\%$ energy shift in some of the features.
The energy calibration in the present work depended primarily on the distance between
the ion cloud and MCP detectors --- this distance was determined with a fractional precision of
1$\%$, with consistent values obtained from both the measurement of the location of the trap electrodes
and detectors and an analysis of the recoil-ion TOF spectra~\cite{Siegl2018}.
This results in a 2$\%$ uncertainty in the neutron energy.
An additional 1\% uncertainty in the neutron energy was attributed to the energy shift applied to
account for the lepton emission.

\subsection{$\beta$n branching ratios}
\label{sec:BDNBranchingRatios}

The $\beta$n branching ratios were obtained by comparing the number of detected
$\beta$-ion coincidences corresponding to decays that emitted a neutron with
energy above 100~keV, $n_{\beta R}$, to the number of detected $\beta$ particles,
$n_{\beta}$, through the relation

\begin{equation}
   P_n = \frac{n_{\beta R}/(\epsilon_{\beta R} \cdot f)}{n_{\beta}/\epsilon_{\beta}},
\label{eq:Pn_1}
\end{equation}
where $\epsilon_{\beta R}$ is the efficiency for detecting the $\beta$-ion coincidences
and $\epsilon_{\beta}$ is the $\beta$-particle detection efficiency.
The ratio $\epsilon_{\beta}/\epsilon_{\beta R}$ was determined from simulations with an
uncertainty of 5$\%$.
The fraction $f$ of the $\beta$n spectrum with neutron energies above 100~keV 
was obtained by assuming the neutron spectrum below threshold has the same intensity
as the 100~keV above threshold, as there is currently no information on the
neutron spectra at these low energies. This assumption yielded values of 0.95(3), 0.89(6),
and 0.83(9) for \textsuperscript{135,136}Sb and \textsuperscript{140}I, respectively, and
an uncertainty of half the difference from unity was assigned.
Given that there do not appear to be pronounced features in the low-energy portions of these spectra,
it is unlikely that there is large peak below 100~keV that would skew these estimates.
To determine $n_{\beta}$, the $\Delta E$ triggers originating from the trapped species of
interest were isolated from those due to decays of isobaric contaminants and other backgrounds. 
This was accomplished by comparing the data to a model that takes into
account the buildup and decay of the different species in the BPT over the course of the
trapping cycle, while enforcing the decay-feeding relationships between the
different populations~\cite{Caldwell2015}. 
The results for $n_{\beta}$ were obtained with 5\% precision.

For $^{135}$Sb and $^{140}$I, the $\beta$n branching ratio was also obtained directly from the
recoil-ion TOF spectrum by comparing $n_{\beta R}$ to the number of $\beta$-ion coincidences
observed for decays without neutron emission, $n_{\beta r}$, using

\begin{equation}
   P_n = \frac{n_{\beta R}/(\epsilon_{\beta R} \cdot f)}{n_{\beta R}/(\epsilon_{\beta R} \cdot f) + n_{\beta r}/\epsilon_{\beta r}}, 
\label{eq:Pn_2}
\end{equation}
where $\epsilon_{\beta r}$ is the efficiency for detecting $\beta$-ion coincidences
without neutron emission and was determined in Ref.~\cite{Munson2018}.
The $n_{\beta r}$ results were obtained by selecting the events in the TOF region
where $\beta$ decays without neutron emission are expected
and subtracting the contribution from isobaric contaminants (when present) and
accidental coincidences.
For $^{135}$Sb and $^{140}$I, $n_{\beta r}$ was determined with 3$\%$ and 6$\%$
precision, respectively. 
The ratio $\epsilon_{\beta R}/\epsilon_{\beta r}$ had an uncertainty of 5$\%$.
For $^{136}$Sb, the trapped \textsuperscript{136}Te activity was substantial enough
to make its subtraction from $n_{\beta r}$ challenging, and therefore, a reliable
value for $n_{\beta r}$ could not be obtained.

\begin{table}
  \caption{\label{tab:BranchingRatiosTable} Recommended $\beta$n branching ratios obtained
  in the present work.  Uncertainties are divided into statistical and systematic.
  } 
\begin{ruledtabular}
\begin{tabular}{ l c c c }
   Isotope & $P_n$ ($\%$) \\
   \hline\noalign{\smallskip}
   \textsuperscript{135}Sb &  14.6 $\pm$ 0.4 (stat) $\pm$ 1.0 (sys)\\
   \textsuperscript{136}Sb & 17.6 $\pm$ 1.0 (stat) $\pm$ 2.6 (sys)\\
   \textsuperscript{140}I & 7.6 $\pm$ 0.9 (stat) $\pm$ 2.6 (sys)\\
\end{tabular} 
\end{ruledtabular}
\end{table}

\begin{figure}[!htbp]
   \subfloat
   {
     \includegraphics[width=0.5\textwidth]{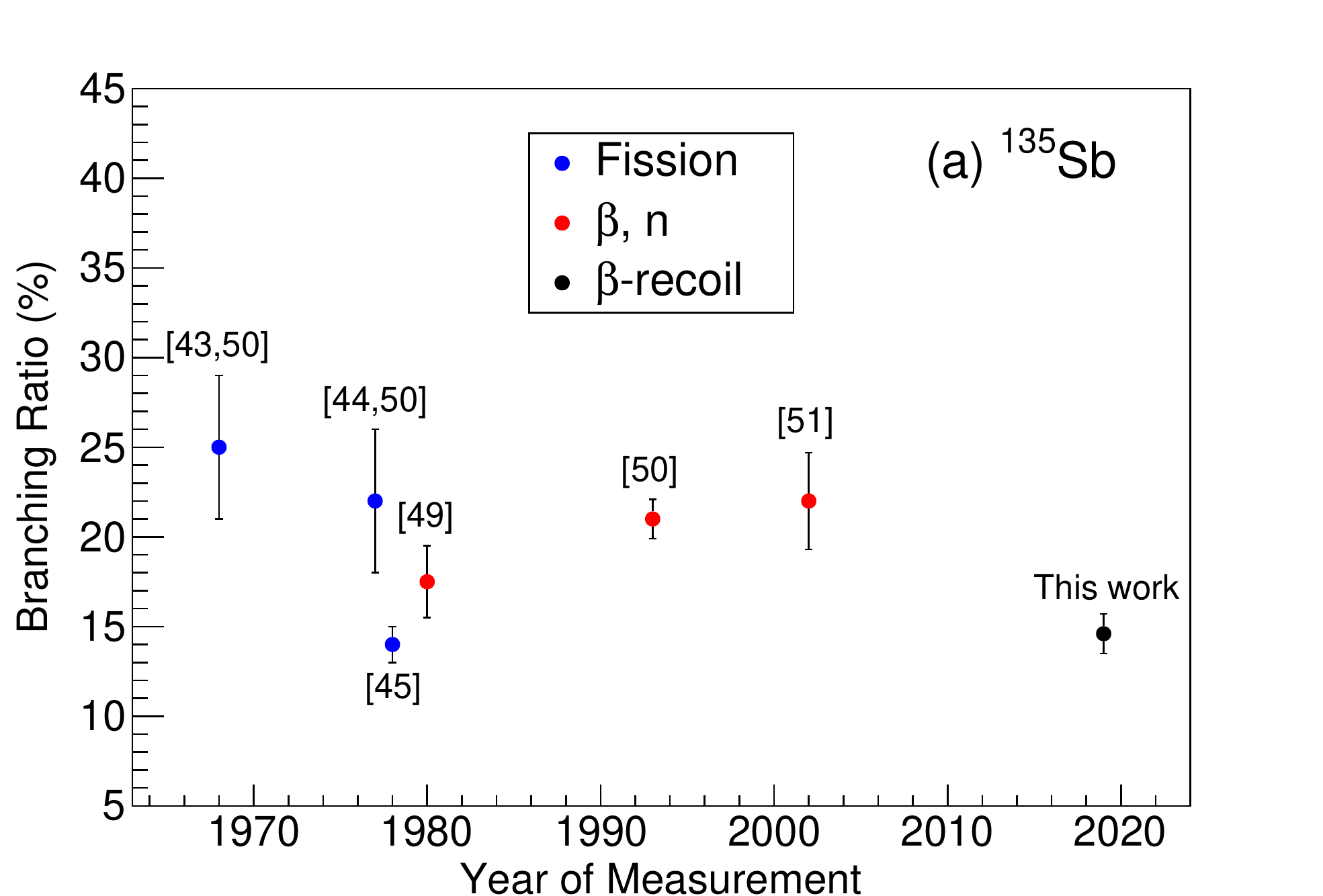}
   }
   \par
   \subfloat
   {
     \includegraphics[width=0.5\textwidth]{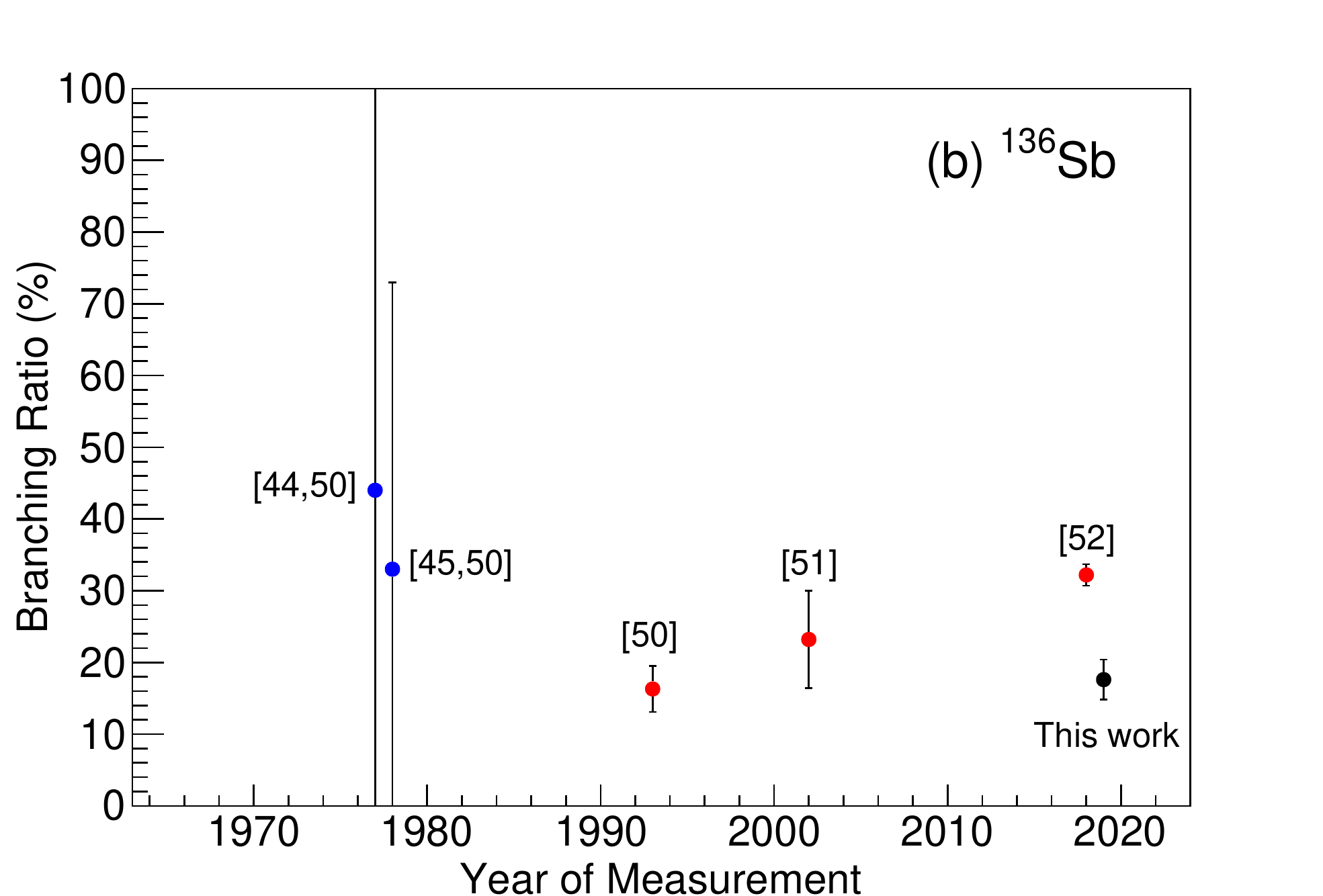}
   }
   \par
   \subfloat
   {
     \includegraphics[width=0.5\textwidth]{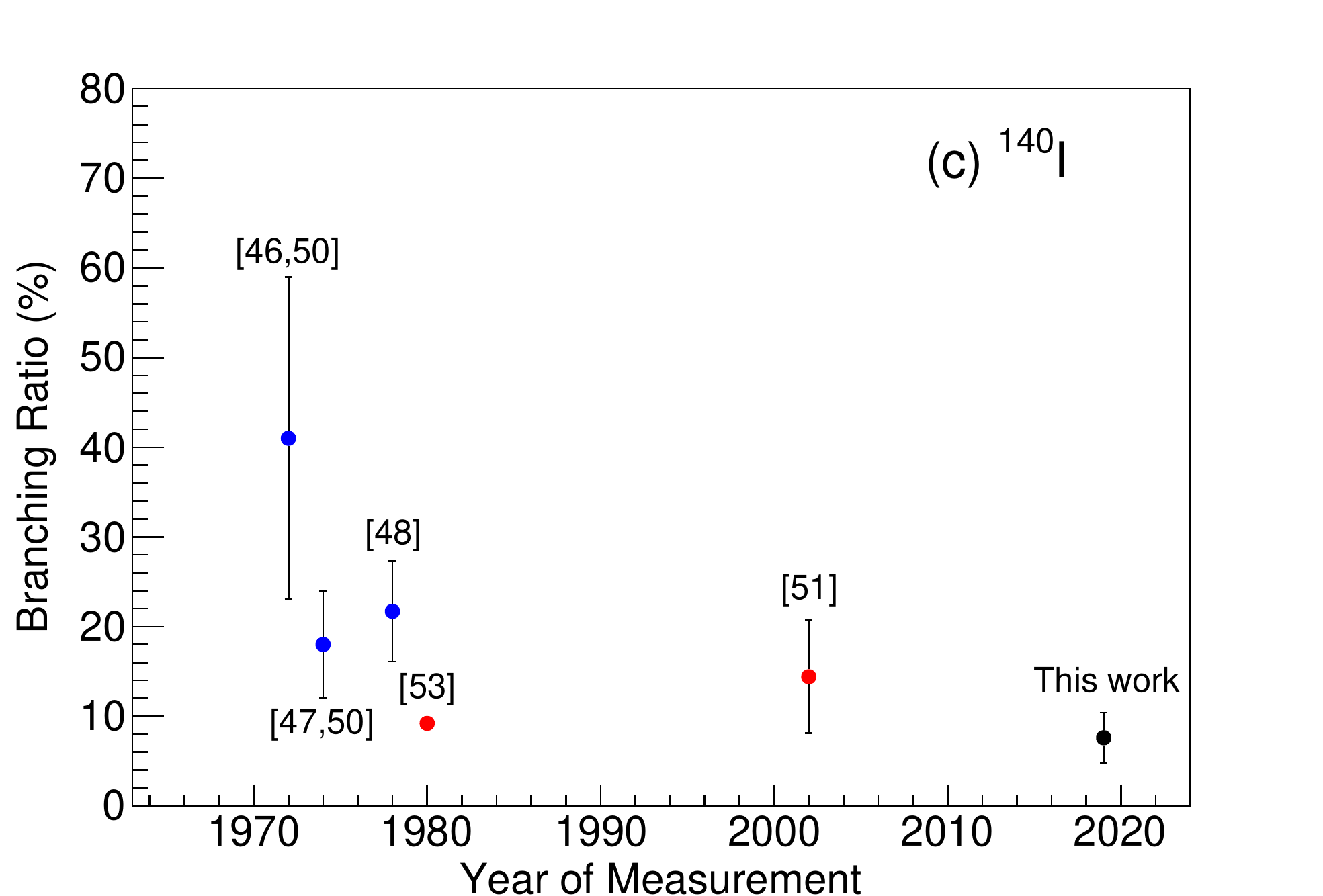}
   }
   \caption
      {
      \label{fig:BranchingRatiosFigure}
      (Color online) Beta-delayed-neutron branching ratios from the present work
      (values taken from Table~\ref{tab:BranchingRatiosTable})
      compared with previous direct measurements for (a)~\textsuperscript{135}Sb,
      (b)~\textsuperscript{136}Sb, and (c)~\textsuperscript{140}I.
      The corresponding year, reference(s), and measurement technique are provided
      for each measurement.
      The label ``fission'' indicates that $P_n$ was obtained from the fission yield and
      neutrons-per-fission of the isotope.
      ``$\beta$, n'' indicates that $P_n$ was obtained by counting $\beta$ particles and neutrons
      separately, usually with plastic scintillators and neutron detectors
      (e.g., BF$_3$ tubes, $^3$He tubes), respectively, and ``$\beta$-recoil'' refers to
      the present work. 
      }
\end{figure}

The $\beta$n branching ratios were obtained from the weighted average of the results from the
four $\Delta E$-MCP detector pairs.
For \textsuperscript{135}Sb and \textsuperscript{140}I,
$P_n$ values of 14.6(16)$\%$ and 8.1(34)$\%$, respectively, were determined from Eq.~\ref{eq:Pn_1},
and values of 14.6(11)$\%$ and 7.6(28)$\%$, respectively, were determined from Eq.~\ref{eq:Pn_2}.  
For \textsuperscript{136}Sb,
Eq.~\ref{eq:Pn_1} yielded a $P_n$ of 17.6(28)$\%$.
In these approaches to determining $P_n$, the systematic uncertainty due to the $\beta$-particle
detection efficiency largely cancels out.  However, obtaining $P_n$ directly from the recoil-ion
TOF spectrum yields a smaller total uncertainty because the systematic uncertainties due to the
MCP solid angles and intrinsic efficiencies also cancel out.
Therefore, for \textsuperscript{135}Sb and \textsuperscript{140}I, the $\beta$n branching ratios
obtained from the recoil-ion TOF spectrum are recommended; for $^{136}$Sb, only the
$P_n$ value obtained from the comparison to detected $\beta$ particles is available.
In Table~\ref{tab:BranchingRatiosTable}, the recommended $\beta$n branching-ratio results
are provided. These values are compared with results obtained from previous direct measurements
in Fig.~\ref{fig:BranchingRatiosFigure}.
In the direct measurements, $P_n$ was determined either from the fission yield and neutrons-per-fission
of the isotope~\cite{Tomlinson1968,Rudolph1977,Crancon1978,Schussler1972,Kratz1974,Kratz1978},
or by counting $\beta$ particles and neutrons
separately~\cite{Lund1980,Rudstam1993,Pfeiffer2002,Caballero-Folch2018,Aleklett1980},
usually with plastic scintillators and neutron detectors (e.g., BF$_3$ tubes, $^3$He tubes),
respectively.
For each isotope, there is roughly a factor of two spread among the $P_n$ results, despite the
fact that in many cases, the quoted uncertainties are significantly smaller than these differences.
These discrepancies are evident even when comparing measurements that used similar experimental
techniques, underscoring the challenging nature of performing $\beta$n spectroscopy and indicating
unforeseen systematic effects were likely responsible for these differences.

The $P_n$ results for $^{135,136}$Sb and $^{140}$I were determined in an analogous manner to the results
for $^{137,138}$I and $^{144,145}$Cs in Ref.~\cite{Czeszumska_Paper}.
In Ref.~\cite{Czeszumska_Paper}, the $\beta$n branching ratios were obtained by comparing the number of
$\beta$-ion coincidences corresponding to $\beta$n decay to the $\beta$-decay activity, which was
measured three different ways: 
(1) from the number of $\beta$ particles detected by the $\Delta E$ detectors,
(2) from the number of $\beta$-ion coincidences registered by the $\Delta E$ and MCP detectors, and
(3) from the number of $\beta$-$\gamma$ coincidences registered by the $\Delta E$ and HPGe detectors.
These three independent measures gave consistent $P_{n}$ results that were in excellent agreement
with previous direct measurements. They also gave an opportunity to probe systematic effects
and provided confidence that they were under control.
In the present work, $P_n$ was obtained using methods (1) and (2),
with limited statistics for $\beta$-delayed $\gamma$-ray emission not allowing method (3).
For $^{135}$Sb and $^{140}$I, where $P_{n}$ from methods (1) and (2) could be compared,
consistent results were again obtained.

\section{Summary and Conclusions}
\label{sec:SummaryAndConclusions}
Beta-delayed-neutron spectroscopy was performed using the BPT instrumented with two plastic-scintillator
$\Delta E$-$E$ telescopes, two MCP detectors, and two HPGe detectors to measure
$\beta$ particles, recoil ions, and $\gamma$ rays, respectively.
Both the $\beta$n energy spectra and branching ratios were determined for the neutron-rich
nuclei \textsuperscript{135,136}Sb and \textsuperscript{140}I.
The $\beta$n energy spectrum for \textsuperscript{136}Sb was measured for the first time, and
the spectra for \textsuperscript{135}Sb and \textsuperscript{140}I were compared
with results from direct neutron measurements by Kratz~\textit{et al.}~\cite{Kratz1979}
and Shalev and Rudstam~\cite{Shalev1977}, respectively.
The $\beta$n energy spectra from the present work were similar in shape and had comparable energy
thresholds to those obtained through direct neutron detection.
The $\beta$n branching ratios were obtained by comparing the number of $\beta$-ion coincidences
from $\beta$n decays to the number of detected $\beta$ decays, which was determined from
the number of $\beta$ particles registered by the $\Delta E$ detector and, when possible,
the number of $\beta$-ion coincidences.
The latter approach to determining the number of detected $\beta$ decays was preferred when available,
as it resulted in smaller systematic uncertainties in $P_n$.

The neutron-energy spectra were obtained with $\beta$-ion-coincidence efficiencies of $\sim$0.5\%, which is
several orders of magnitude larger than the neutron-detection efficiencies achievable with the $^{3}$He and
gas-proportional detectors used for direct neutron spectroscopy.
The ion-trap approach is therefore well suited for use at radioactive-beam facilities, where efficient
techniques are desired to make the most of the delivered beam intensities. 
The $\beta$n branching ratios for $^{136}$Sb and $^{140}$I were determined with beam intensities
of only 5~ions/s, and with improvements to the detector array, results could be obtained with
beams of less than 1~ion/s.

Upgrades to the BPT setup are currently in development. Plans include increasing the
$\beta$-recoil-coincidence detection efficiency using larger plastic scintillators and MCP detectors,
and lowering the neutron energy threshold by further minimizing the impact of 
the electric fields on the trajectories of the recoil ions.
The latter will be accomplished by bringing the electrodes closer to the center of the ion trap to
allow for a lower-amplitude RF voltage to be applied, thus reducing the perturbation of the ion
trajectories while the ions are in transit to the MCP detectors. 
Future experiments will also benefit from the increased intensities and purities of the beams delivered
by the CARIBU facility \cite{Hirsh2016,Savard2016}; since these measurements were performed,
the beam intensities have increased by an order of magnitude.  These improvements will allow
$\beta$n measurements to be performed for neutron-rich nuclei even further from stability,
allowing access to many of the isotopes that significantly impact \textit{r}-process nucleosynthesis.

\section{Acknowledgements}
We acknowledge and appreciate the assistance of the
ATLAS staff. This material is based upon work supported
by the Department of Energy, National Nuclear Security
Administration, under Award Numbers DE-NA0000979 (NSSC),
DE-AC52-07NA27344 (LLNL), and DE-NA0002135 (SSGF);
Office of Nuclear Physics Contract DE-AC02-06CH11357 (ANL);
NEUP Project Number 13-5485 (University of California);
Grant DE-FG02-94ER40834 (University of Maryland);
Louisiana State Board of Regents Research Competitiveness Subprogram LEQSF(2016-19)-RD-A-09;
NSERC, Canada, under Application No. 216974;
NSF contract PHY-1419765;
and the Department of Homeland Security.

\bibliography{Bibliography}

\end{document}